\documentclass[doublecol]{epl2} 
\usepackage[colorlinks=true,
            linkcolor=blue,
            citecolor=blue,
            urlcolor=blue]{hyperref}

\usepackage{amsmath,amssymb}
            
\title{Time-Symmetry of Lagrangian Coherent Structures in Active Turbulence}
\shorttitle{Time-Symmetry of Lagrangian Coherent Structures in Active Turbulence} %Insert here a short version of the title if it exceeds 70 characters

\newcommand{\sif}{\sigma_F}
\newcommand{\sib}{\sigma_B}

\author{Suvarchalanjan Bellaganti\inst{1}\thanks{These authors contributed equally to this work.} \and Amal Manoharan\inst{1,2}$^{\rm{(a)}}$ \and Kirti Kashyap\inst{3} \and Siddhartha Mukherjee\inst{1}\thanks{E-mail: \email{smukherjee@iitk.ac.in} (Corresponding Author)}
}
\shortauthor{S. Bellaganti\etal}

\institute{                    
  \inst{1} Department of Mechanical Engineering, Indian Institute of Technology Kanpur, Kanpur 208026, India\\
  \inst{2} International Centre for Theoretical Sciences, Tata Institute of Fundamental Research, Bengaluru 560089, India\\
   \inst{3} Department of Physics, Indian Institute of Technology Hyderabad, Hyderabad, India
}

\abstract{Active flows are central to mixing and transport across living systems. While Newtonian fluids remain laminar, diffusive and predictable at the microscale, living fluids like dense bacterial suspensions can exhibit highly chaotic flows like \textit{active turbulence}, with anomalous transport capabilities. The underlying spatiotemporally persistent structures that drive mixing in active flows, however, remain uncharted. Using Lagrangian Coherent Structures, we now uncover networks of attracting and repelling hyperbolic surfaces. We study changes in the distribution and spectra of Finite-Time Lyapunov Exponent fields in response to increasing activity. Despite the dominance of vorticity in the flow, extreme forward and backward time chaotic mixing is found to originate from straining regions, emphasizing the role of saddles. Fractal dimensions of ridges reveal a morphological simplification of LCS networks with increasing activity, while retaining isotropic crossing. Throughout our work, we also probe a hitherto unasked question---Are signatures of Lagrangian irreversibility manifest in attracting and repelling LCS? To the contrary, we find there is a striking time-symmetry. Our work takes the first steps towards linking flow structures in active turbulence to invariant mixing surfaces. These findings will crucially help in designing activity modulation protocols to seed or inhibit flow structures, and thence mixing, in a bid to tame active turbulence for varied applications.}

\begin{document}

\maketitle

\section{Introduction}
Active flows, from microscopic intra-cellular motion to mesoscopic collective behaviour of cells, are essential to vital biological processes like growth, morphogenesis, swarming and bioconvection, shaping transport and mixing across living systems~\cite{verstraeten2008living,hill2005bioconvection,marchetti2013hydrodynamics,vafa2022active,hoffmann2022theory,narayanasamy2025metabolically,plum2025morphogen}. While microscale flows in Newtonian fluids are limited to laminar states with weak, diffusion driven mixing, active flows surprisingly overcome these bounds by exhibiting highly complex flowing states like \textit{active turbulence}~\cite{wensink2012meso,dunkel2013fluid,alert2022active}, enabling enhanced material transport. Studies have since explored the analogy of active with inertial (high Reynolds) turbulence. Active turbulence has been found to exhibit diverse phenomenology, with transitions from non-universal~\cite{bratanov2015new} to universal scaling states beyond a critical degree of activity $\alpha_c$ accompanied by the emergence of intermittency~\cite{mukherjee2023intermittency,kiran2024onset}. These states are accompanied by crucial changes in the underlying flow, reflected in the emergence of novel features like vorticity streaks, local ordering and spatial heterogeneity~\cite{mukherjee2021anomalous,puggioni2023flocking,kashyap2025}, which do not have analogues in inertial turbulence. Activity heterogeneities, moreover, give rise to fluctuating hydrodynamic interfaces, coexisting flow states and growing fronts~\cite{mukherjee2025,sahoo2026hydrodynamics}, akin to turbulent/non-turbulent interfaces in inertial turbulence. 

The Lagrangian picture is equally rich, a key finding being that beyond a critical activity, active turbulence displays robust anomalous diffusion~\cite{mukherjee2021anomalous,singh2022lagrangian}, which has been experimentally observed as a cell-density driven effect~\cite{gautam2024harnessing}. Mean square displacements (MSDs) of tracer particles are found to be superdiffusive, mediated via L\'evy walks~\cite{ariel2015swarming,mukherjee2021anomalous,gautam2024harnessing}. A closer look at the hydrodynamics has linked changes in turbulence organization to preferential ballistic motion, due to emergent dynamical heterogeneity~\cite{singh2022lagrangian}.

\begin{figure*}
\includegraphics[width=\linewidth]{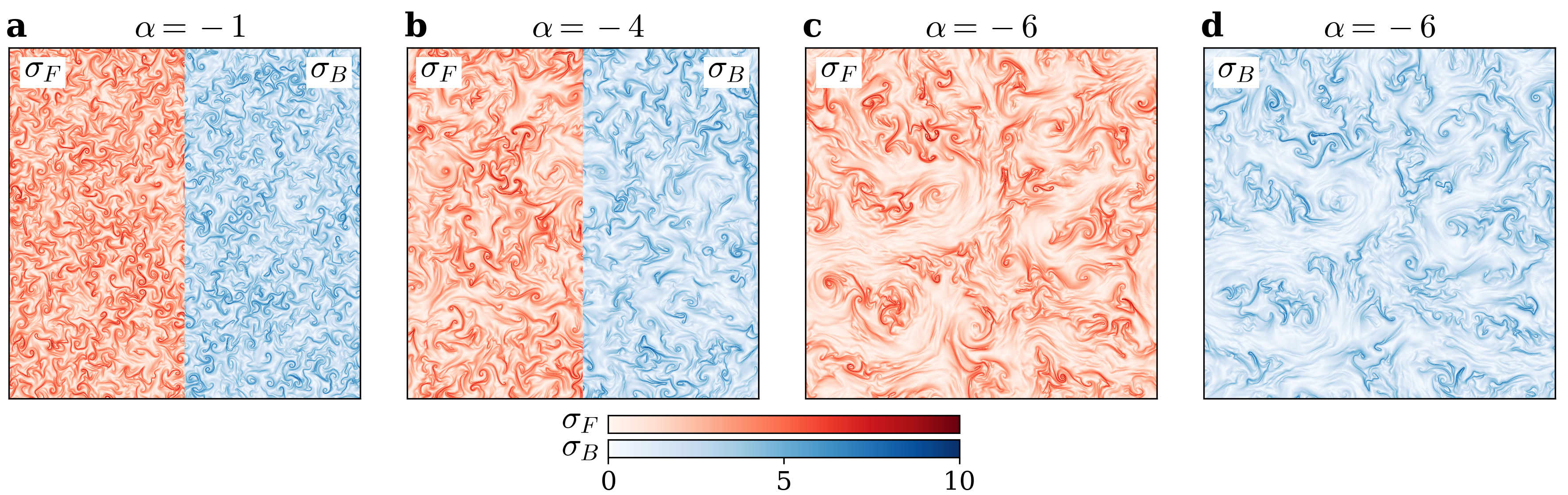}
\caption{\textbf{FTLE Fields.} FTLE fields for increasing activity are shown for an integration time $\mathcal{T} = 0.5$. In \textbf{(a)} and \textbf{(b)}, the left and right halves show the $\sigma_F$ and $\sigma_B$ fields. As activity increases, the FTLE fields change from being densely packed with small fronts of a single lengthscale (as seen in \textbf{(a)} $\alpha = -1$) to a more heterogeneous organization with sparse islands separating clusters of high FTLE in \textbf{(b)}. This is most prominently seen in \textbf{(c)} $\sigma_F$ and \textbf{(d)} $\sigma_B$, for $\alpha = -6$. The FTLE fronts tend to elongate, while the fields become more multiscale and heterogeneous, with a segregation of front clusters and calm patches.}
\label{fig1}
\end{figure*}

Another perspective on mixing in active turbulence comes through studies of chaos. Lyapunov exponents ($\lambda$), measuring exponential divergence of initially close solution trajectories in phase-space (Eulerian chaos), are found to increase with activity~\cite{mukherjee2023intermittency}, while remaining bounded due to a lack of scale separation. This partially echoes inertial turbulence where $\lambda$ exhibits an unbounded power-law growth with Reynolds number (Re)~\cite{mukherjee2016predictability,boffetta2017chaos,banerjee2025intermittent}. Lyapunov spectra and dimension for active turbulence have also been recently quantified~\cite{boutros2026global}. There is a caveat, however, that turbulence organization comprises multiple scales of motion, while chaos is globally dominated by the \textit{maximal} Lyapunov exponent linked to the smallest, hence fastest, scale of motion~\cite{lorenz1969predictability,rotunno2008generalization,palmer2014real,mukherjee2016predictability}. These exponents are also meaningful only in the limit of infinitesimal perturbations tracked asymptotically~\cite{cencini2013finite}, which fails to capture the vivid heterogeneity of mixing across space, scale and Lagrangian histories. With burgeoning interest in designing tunable active flows, it is crucial to study chaotic mixing through a lens that reveals its persistent spatiotemporal structures.

We do this by uncovering Lagrangian Coherent Structures (LCS)~\cite{haller2000lagrangian,mathur2007uncovering,haller2011lagrangian,peacock2013lagrangian,haller2015lagrangian} in active turbulence, which form the interconnected networks of attracting and repelling fronts associated with chaotic stretching of finite perturbations over finite time. In living systems, LCS drive morphogenesis~\cite{serra2020dynamic} and are altered by activity, affecting mixing in cellular flows~\cite{serra2020dynamic,ran2021bacteria,serra2023defect,si2024interaction}. As a first study of LCS in active turbulence, we quantify their statistics and organization with varying activity, and establish crucial links between instantaneous Eulerian flow structures with forward and backward time Lagrangian mixing. In the light of Lagrangian time-asymmetry in turbulence, as evidenced in irreversibility~\cite{xu2014flight,picardo2020lagrangian,kiran2023} and pair-dispersion~\cite{shnapp2025velocity}, we ask whether there is any signature of time-asymmetry detectable in LCS. We extend the study of LCS network geometry with fractal dimensions of ridges and crossing angles. Overall, we show how activity decisively alters chaotic mixing patterns, while finding that forward and backward time LCS are remarkably similar. Augmenting ongoing work on exploiting activity heterogeneities to control flow structures, our work will aid in tuning mixing in active flows.

\section{Methods}
We solve the incompressible ($\nabla \cdot {\bf u}$) Toner-Tu Swift-Hohenberg equation~\cite{wensink2012meso}, to model bacterial turbulence, where the velocity field $\mathbf{u}$ evolves as:
\begin{equation}
\partial_t{\bf u} + \lambda_0 {\bf u}\cdot {\nabla\bf u} =-{\bf \nabla}p-\Gamma_0\nabla^2{\bf u}-\Gamma_2\nabla^4{\bf u}-(\alpha + \beta|{\bf u}|^2){\bf u},
\label{GNS}
\end{equation}
The parameter $\lambda_0 >1$ corresponds to swimmer type (pushers) and the $\Gamma$-terms lead to lengthscale selection. The last term is the Toner-Tu drive where $\alpha < 0$ (activity parameter) leads to local polar ordering of the fluid and $\beta>0$ for stability. We perform pseudospectral direct numerical simulations of Eq.~\ref{GNS} in 2D, using a $N/4$ de-aliased to account for the cubic non-linearity, on square periodic boxes of length $L=20$ discretized over $N^2=1024^2$ collocation points, with a semi-implicit Crank-Nicholson scheme with time-stepping $\Delta t = 0.0002$. All parameters are kept the same as in earlier works, i.e. $\Gamma_0 = 0.045, \Gamma_2 = \Gamma_0^3, \beta = 0.5, \lambda = 3.5$~\cite{wensink2012meso,james2018turbulence,james2018vortex,cp2020friction,mukherjee2021anomalous,singh2022lagrangian,mukherjee2023intermittency}. 

\subsection{Finite-Time Lyapunov Exponents}
\begin{figure*}
\includegraphics[width=\linewidth]{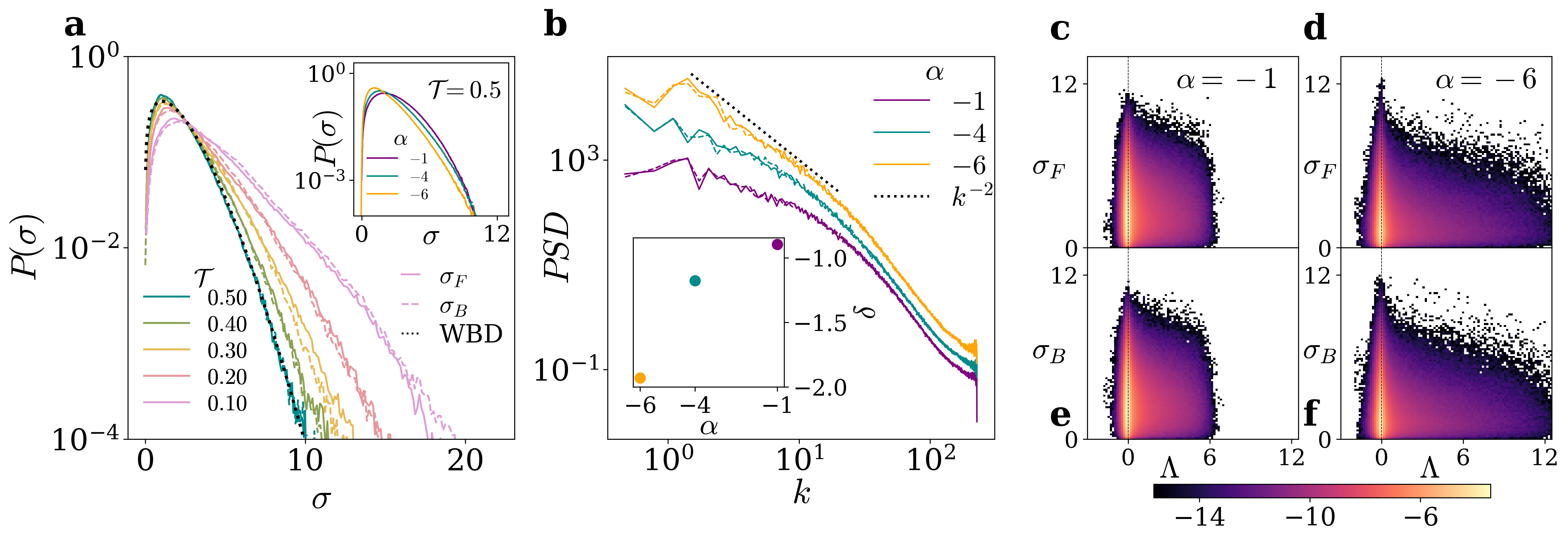}
\caption{\textbf{Statistics of FTLE Fields}. \textbf{(a)} Probability distributions of $\sif$ and $\sib$ (for $\alpha = -6$) tend to contract for increasing integration times $\mathcal{T}$, while the tail is approximately exponential, specially for low $\mathcal{T}$ (broken black line shows a Weibull distribution fit). Moreover, $\sif$ and $\sib$ distributions coincide, showing that the forward and backward FTLE fields have identically distributed values. (Inset) Distribution of $\sigma$ for different $\alpha$. \textbf{(b)} Power-spectra of the FTLE fields for increasing activity (vertically staggered for clarity). The spectra become steeper, tending towards $k^{-2}$ as $\alpha \lesssim \alpha_c \approx -5$, with a slight increase in the range of wavenumbers over which power-law scaling is observed. This shows that the FTLE field is more hierarchical with a dominance of large scale structures over the small scales. (Inset) Scaling exponents $\delta$ from power-law fits to the low wavenumber part of the spectra which goes as $k^\delta$. \textbf{(c)}-\textbf{(f)} Joint-pdfs of the $\sif$ and $\sib$ fields with the Okubo-Weiss $\Lambda$ field (at $t_0$), for different activity (colobar is logarithmic). As activity increases the range of $\Lambda$ increases significantly in the $\Lambda \gg 0$ direction showing that vorticity is much more dominant than strain in active turbulence. There is also a slight increase in the range of $\sif$ and $\sib$ with increasing activity. However, even though the $\Lambda$ field is highly skewed towards a dominance of vorticity ($\Lambda \gg 0$), as seen in \textbf{(c)} and \textbf{(d)} for $\alpha = -6$, the strongest mixing both in forward and backward time ($\sif \gg 1$ and $\sib \gg 1$) preferentially originates in the strain dominated regions of the flow ($\Lambda < 0$).}
\label{fig2}
\end{figure*}

With the velocity field $\mathbf{u}(\mathbf{x},t)$, we associate the flow map 
\begin{equation}
\mathbf{F}_{t_0}^{t}(\mathbf{x}_0) = \mathbf{x}_0 + \int_{t_0}^t \mathbf{u}( \mathbf{F}_{t_0}^\tau (\mathbf{x}_0), \tau) \mathrm{d}\tau
\end{equation}
which maps an initial particle position $\mathbf{x}_0$ at time $t_0$ to its position at time $t$, giving the deformation gradient tensor $\nabla \mathbf{F}_{t_0}^{t}(\mathbf{x}_0)$. The right Cauchy--Green strain tensor is obtained as $\mathbf{C}_{t_0}^{t}(\mathbf{x}_0) = \nabla \mathbf{F}_{t_0}^{t}(\mathbf{x}_0)^{\rm T} \nabla \mathbf{F}_{t_0}^{t}(\mathbf{x}_0)$, which is symmetric and positive definite, and has eigenvalues $\lambda_1 \leq \lambda_2$ and associated eigenvectors $\lbrace \boldsymbol{\xi}_1,\boldsymbol{\xi}_2 \rbrace$. The finite-time Lyapunov exponent (FTLE) field is then defined by
\begin{equation}
\sigma_F(\mathbf{x}_0) := \sigma_{t_0}^{t}(\mathbf{x}_0) =
\frac{1}{\mathcal{T}}
\ln \sqrt{\lambda_2(\mathbf{x}_0)}
\end{equation}
where $\lambda_2$ denotes the largest eigenvalue of $\mathbf{C}_{t_0}^{t}(\mathbf{x}_0)$ and $\mathcal{T} = |t - t_0|$ is the integration time. Forward time integration gives the $\sif$ (Forward FTLE) field, while backward integration gives $\sib$ (Backward FTLE). Numerically, the flow map is obtained by advecting $N\times N$ tracer particles over the finite integration interval $\mathcal{T}$. The deformation gradient tensor is subsequently evaluated using finite differences on the flow map, after which $\mathbf{C}_{t_0}^{t}(\mathbf{x}_0)$ and its eigenvalues are computed to obtain the FTLE fields~\cite{haller2015lagrangian}.

\section{Results}
We begin with a visual impression in Fig.~\ref{fig1} of the forward ($\sigma_F$) and backward ($\sigma_B$) FTLE fields with varying activity $\alpha$ and $\mathcal{T} = 0.5$ (which corresponds to roughly $10\tau_\omega$, where $\tau_\omega \sim 1/\omega^\prime$ with $\omega^\prime$ the rms vorticity). As activity increases, it has been observed that the flow becomes multiscale, transitioning to a universal power-law spectrum when $\alpha \lesssim \alpha_c \approx -5$~\cite{mukherjee2023intermittency}, while the vorticity field becomes inherently heterogeneous with the emergence of streaks, large coherent vortices and local ordering~\cite{mukherjee2021anomalous,singh2022lagrangian,kashyap2025,perlekar2026flocking}. Similar features are reflected in the changes observed in the FTLE fields with increasing activity. At low activity (Fig.~\ref{fig1}a), where the flow is mildly turbulent and populated by densely packed vortices of a single lengthscale, the $\sigma_F$ (left) and $\sigma_B$ (right) fields also consist of densely packed fronts of roughly the same size across the flow. The mushroom-shaped fronts of the FTLE fields resemble counter-rotating vortex pairs (but are not trivially associated with them). The $\sigma_B$ field exhibits identical, but not coincident, features as the $\sigma_F$ field. The fronts intensify with increasing activity, become elongated and sparse, with signs of local ordering. This is most pronounced at $\alpha = -6$ (Fig.~\ref{fig1}c). Large calm patches with mild $\sigma_F$ tendrils separate intensely knotted clusters of strong $\sigma_F$ fronts. This is also consistent with the Lagrangian picture, where at high levels of activity there are flow regions where tracer packets move coherently and ballistically without deforming~\cite{singh2022lagrangian}, which will naturally lead to lower $\sigma_F$ values.

It appears that the forward and backward FTLE fields both have a very similar distribution of structures across activity, as seen from comparing the left and right halves of Fig.\ref{fig1}a-b. Fig.~\ref{fig1}d shows the full $\sigma_B$ field starting from the same initial configuration as the $\sigma_F$ field in Fig.~\ref{fig1}c. Both fields appear very similar in terms of their structure geometry, distribution and relative volume fractions of fronts to empty regions. Naturally, the fields are not coincident and are pointwise different, which can be readily confirmed from their joint distribution yielding no correlation (not shown). At a first glance, this seems to imply forward and backward LCS are symmetric in active turbulence. To the best of our knowledge, there has not been a comparison between the geometry of forward and backward LCS even for inertial turbulence. \textit{A priori}, this is difficult to rationalize, given that both active and inertial turbulence are marked by distinct Lagrangian time-asymmetries~\cite{xu2014flight,shnapp2025velocity,picardo2020lagrangian,kiran2023}, and backward mixing is found to be faster than forward mixing in inertial turbulence~\cite{sawford2005comparison,berg2006backwards,salazar2009two,bragg2016forward,shnapp2025velocity}. As such, one may expect manifestations of these effects in the LCS. We check for this while quantifying the statistics and geometry of forward vis-\`a-vis backward FTLE fields and LCS.

\begin{figure*}
\centering
\includegraphics[width=\linewidth]{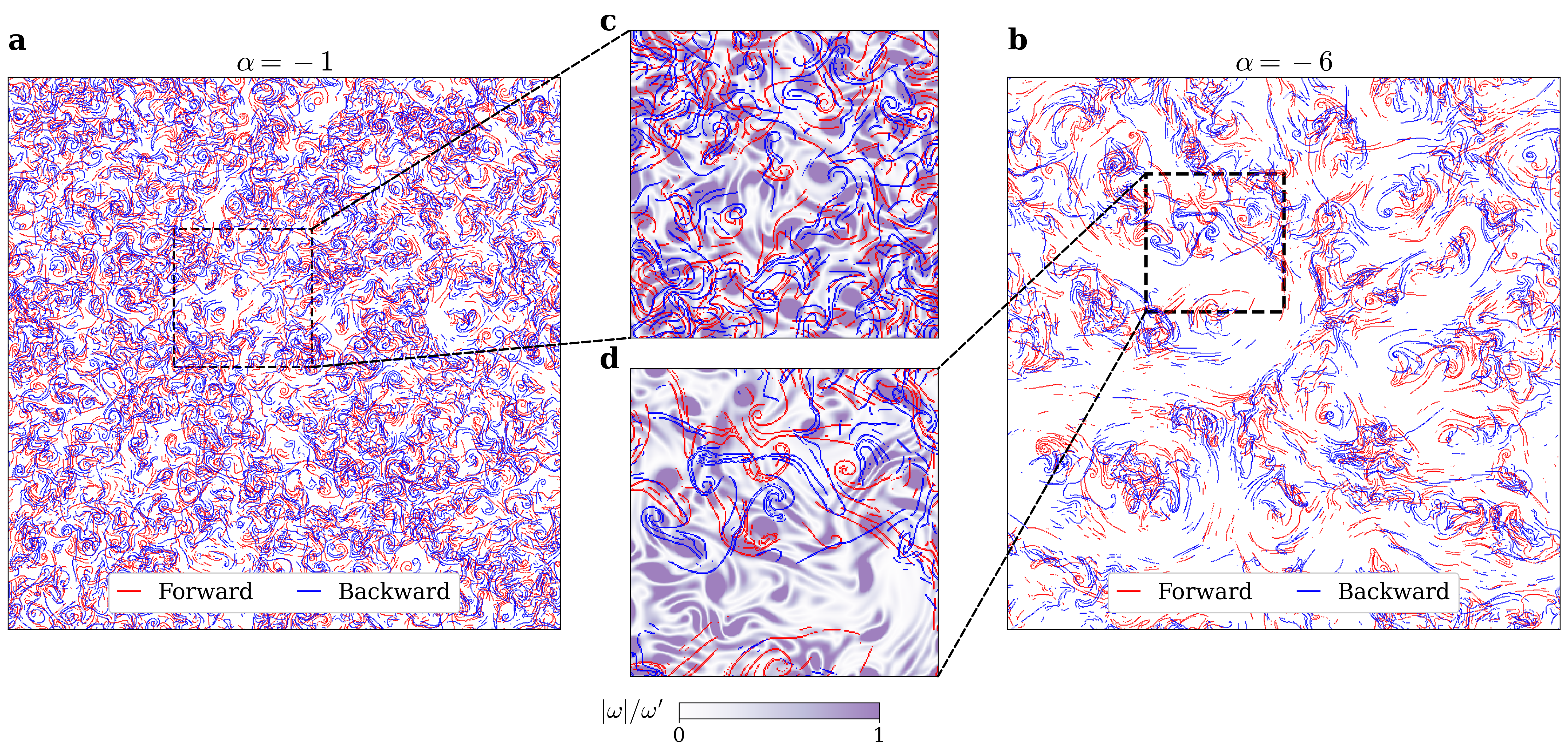}
\caption{\textbf{Lagrangian Coherent Structures in Active Turbulence.} Ridges from the forward (red) and backward (blue) FTLE fields that form the network of Lagrangian Coherent Structures have been shown for \textbf{(a)} $\alpha = -1$ and \textbf{(b)} $\alpha = -6$. As the flow organization becomes more heterogeneous at higher activity, the LCS networks also open up into empty patches separated by dense clusters of interlocked ridges. Magnified sections in \textbf{(c)} and \textbf{(d)} further show that the dense ridge regions also become less populated at higher activity. The initial time vorticity magnitude field has been superposed in the background, which shows that the LCS structures are not trivially correlated with vortices.}
\label{fig3}
\end{figure*}

\subsection{Statistics of FTLE fields} We compare the probability distribution of FTLE amplitudes in Fig.~\ref{fig2}a, for the highest activity $\alpha=-6$. As $\mathcal{T}$ increases, extreme short-time mixing gets averaged out as particles sample multiple flow regions, which reduces the variance of the pdfs, while they are found to closely follow Weibull distributions. The pdfs of $\sigma_F$ (solid lines) are found to coincide with $\sigma_B$ (dashed lines), reflecting that at the level of amplitudes, the forward and backward fields are identically distributed. The inset shows the pdfs for different $\alpha$, and a fixed $\mathcal{T} = 0.5$. There is a slight reduction in the mean $\sigma$ values with increasing activity. While the amplitude distribution reflects a symmetry between forward and backward mixing, there may be variation in the details of spatial organization of structures and their distribution. 

We probe this in Fig.\ref{fig2}b with the shell-averaged power spectral density (PSD) of the $\sigma_F$ and $\sigma_B$ fields over wavenumber $k$ (the spectra have been vertically staggered for clarity). The effects noticed in the FTLE fields upon increasing activity in Fig.\ref{fig1} are now quantified. At low activity ($\alpha = -1$), the spectrum has a shallow slope over a small range of wavenumbers. This means that small scale structures (higher $k$) are not entirely dominated by large structures (low $k$), as the ratio of their energies is small. As activity increases, the slope of the spectra changes from approximately $k^{-1}$ to $k^{-2}$ (inset shows the exact values of the scaling $k^\delta$, calculated from power-law fits). The slope increasing to $\approx k^{-2}$ as $\alpha \lesssim \alpha_c \approx -5$ reflects a stronger hierarchy in the distribution of structures, where the pronounced peak at low $k$ shows a dominance of larger structures in the FTLE fields. Interestingly, there is again a strong coincidence between the forward and backward FTLE spectra, which extends the similitude of these fields from their amplitude distributions to their spatial organization as well. Therefore, at this level, forward and backward FTLE fields in active turbulence are structurally the same. This implies that mixing proceeds with the same spatial heterogeneity, rate, and has very similar patterns both forward and backward in time, when measured via FTLEs. Of course, the fact that the $\sif$ and $\sib$ are not coincident with each other encodes the pointwise temporal asymmetry in mixing. 

An important question then is how do flow structures influence the emergence of FTLE fronts? In two-dimensional turbulence, the usual marker for local flow topology is the Okubo-Weiss parameter $\Lambda = (\omega^2 - s^2)/4\langle \omega^2 \rangle$ which distinguishes between the local dominance of strain over vorticity. Here $s = (\partial_j u_i + \partial_i u_j )/2$ and $\omega = (\partial_j u_i - \partial_i u_j )$. Regions of the flow with $\Lambda > 0$ represent vorticity cores while these are usually surrounded with a halo of strain ($\Lambda < 0$)~\cite{singh2022lagrangian}. How these flow structures relate to the FTLE fields is best quantified by looking at the joint distribution of $\sif$ and $\sib$ with $\Lambda$. There is an important caveat here that $\sigma$ fields encode \textit{spatiotemporal} information while, on the other hand, the $\Lambda$ field is instantaneously determined by the strain and vorticity fields, and hence is purely Eulerian. One can ask, however, that starting from a particular flow region as identified by $\Lambda$, what is the most likely mixing outcome as quantified by $\sif$ and $\sib$? 

Fig.~\ref{fig2}c-\ref{fig2}f show the joint distributions of $\sif$ and $\sib$ with $\Lambda$ (for $\mathcal{T} = 0.5$). We recall first that $\Lambda$ has a distribution highly skewed towards $\Lambda>0$, which reflects that vorticity attains much higher values than strain~\cite{singh2022lagrangian} (we measure the skewness going from approximately $4$ to $7$ as $\alpha$ goes from $-1$ to $-6$). The joint-pdfs show that strain dominates early time stretching and mixing of tracers, as higher values of $\sif$ are preferentially found in regions of high strain (i.e. in the $\Lambda <0$ part of the distribution) for all $\alpha$ values, even though $\Lambda>0$ regions are much more populous. The main effect of increasing activity is that the peak corresponding to the highest possible value of $\sif$ increases as $\alpha \lesssim \alpha_c$, and $\Lambda \gg 0$ becomes prevalent. Curiously, even in active turbulence where vorticity dominates (reflected in the skewness of $\Lambda$), mixing occurs preferentially starting from strain regions. This can be understood as strain dominated regions locally behaving like saddle-points that are crucial for chaotic stretching and folding, while vortices mediate the reinjection into these regions, as readily observed in simpler models like vortex arrays~\cite{witt1999tracer}. Moreover, in highly active turbulence, regions emerge where the local streamline structure is induced dominantly by the large scale, polar-ordered vortices~\cite{mukherjee2021anomalous,singh2022lagrangian,puggioni2023flocking,kashyap2025}. This makes trajectories persistent, akin to the ``sweeping effect'' in inertial turbulence. All this is reflected in the joint distribution showing a low probability for attaining high FTLE values ($\sif,\sib \gg 1$) starting from vorticity dominated regions $(\Lambda \gg 1)$. 

The joint distribution of $\sib$ with $\Lambda$ is again identical to the distribution of $\sif$. This shows that the forward and backward time dynamics of mixing is identical at the level of LCS in active turbulence. Starting in high strain leading to stronger forward and backward time mixing is the more likely outcome, while strong mixing starting from vortical regions is comparatively rare.

\subsection{Geometry of Lagrangian Coherent Structures} 
We now turn to the ridge geometry of the FTLE fields. Ridges are surfaces (or lines, in two-dimensions) in the $\sif$ and $\sib$ fields, which mark strong inflection in the Lyapunov exponent values transverse to the surface. These are found from the negative eigenvalue of the local Hessian of the FTLE fields~\cite{mathur2007uncovering}. The interconnected network of these ridges forms the skeleton of turbulent mixing, and these are referred to as Lagrangian Coherent Structures~\cite{haller2015lagrangian}. In Fig.~\ref{fig3} we show the LCS for active turbulence with different levels of activity, where the forward (repelling) LCS have been shown in red and backward (attracting) LCS in blue, as conventional. In simple terms, these ridges mark regions which lead to most intense mixing, forward and backward in time. Fig.~\ref{fig3}a shows the LCS for mildly active turbulence with $\alpha = -1$, which reveals a dense network of tightly packed ridges. These knotted structures essentially form over a small range of lengthscales, comparable to the dominant vortex lengthscale in the flow. Fig.~\ref{fig3}b shows the LCS for highly active turbulence with $\alpha = -6$. The ridges now form a heterogeneous distribution, with large voids (calmer mixing zones) separating islands of interlocked LCS. The length of the ridges appears to increase with activity, as longer filaments are formed. Fig.\ref{fig3}c and \ref{fig3}d show magnified sections of the LCS for $\alpha = -1$ and $-6$, respectively, superimposed over the initial time vorticity field magnitude. While there are some signatures of the LCS avoiding the \textit{streak} regions of the vorticity field for $\alpha=-6$, very clear parallels between the spatial organization of vorticity and LCS is not obvious. This highlights the importance of studying the spatiotemporal structures of mixing via LCS in its own right. 

\begin{figure}
\centering
\includegraphics[width=\linewidth]{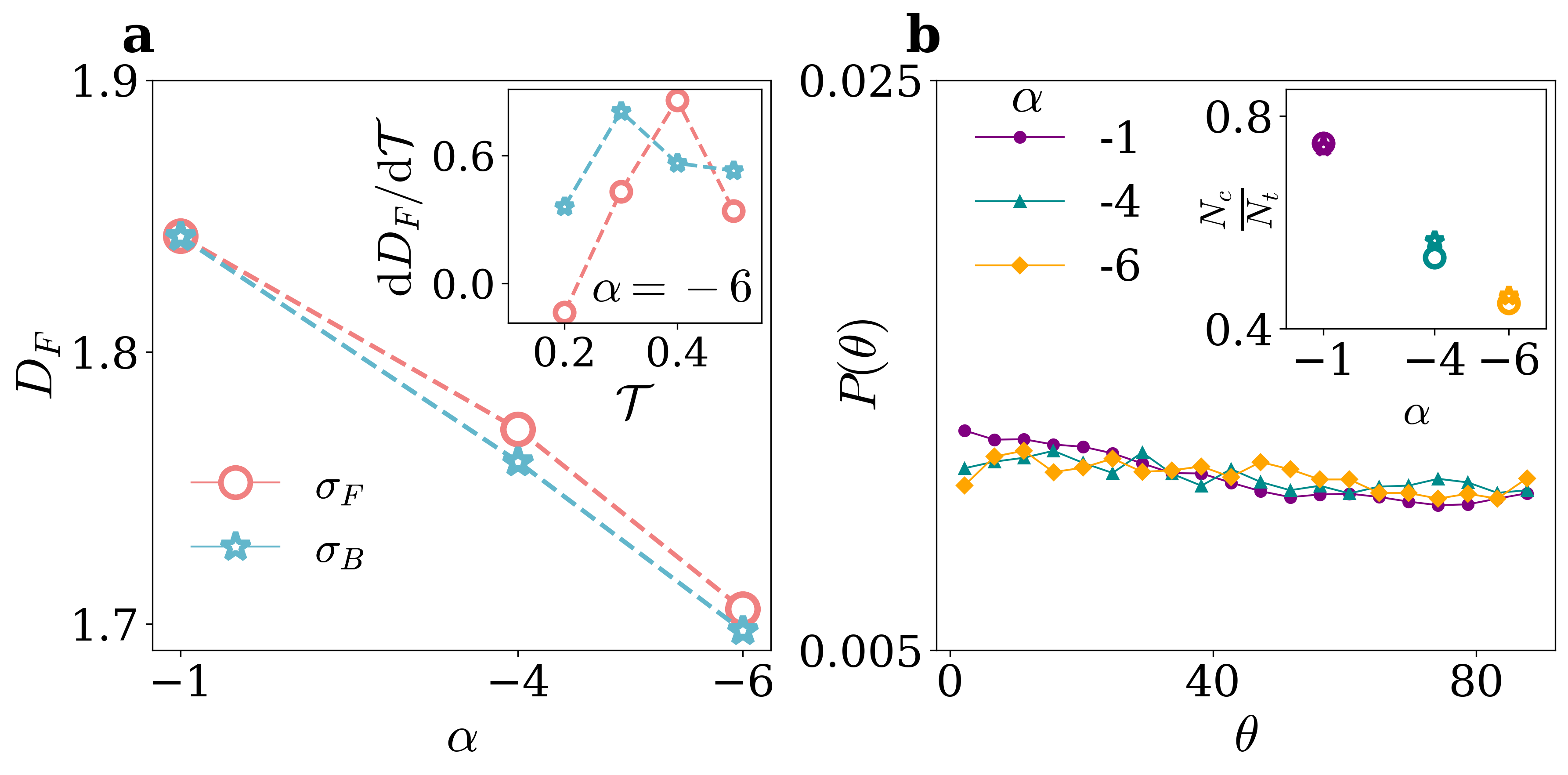}
\caption{\textbf{LCS Geometry.} \textbf{(a)} Fractal dimension $D_F$ of the ridges shows that attracting and repelling LCS have the same geometrical features, as they yield the same $D_F$ values. As activity increases, the ridges tend to elongate into more linear shapes which causes a dip in $D_F$. Inset shows the change in the slope $\mathrm{d}D_F/\mathrm{d}\mathcal{T}$ with $\mathcal{T}$, which after an initial increase begins to dip, hinting that the morphology converges over $\mathcal{T}$. \textbf{(b)} PDFs of the intersection angle $\theta$ of attracting and repelling LCS is found to be uniform. There is no preferential orthogonality between the forward and backward time LCS, and their crossings are found to be isotropic. Inset shows the density of crossings, $N_c/N_t$, decreases with activity.}
\label{fig4}
\end{figure}

To quantify the geometry of ridges, we compute their fractal dimension $D_F$ which serves as a measure of how convoluted the LCS are (where straight lines would yield $D_F = 1$ and space filling fractal curves would give $D_F \approx 2$). We show $D_F$ obtained via box-counting, for forward and backward LCS, and for varying activity in Fig.~\ref{fig4}a. There is a clear reduction in $D_F$ as the activity increases, which reflects the lengthening of the LCS as $\alpha \lesssim \alpha_c$. Even so, the absolute value of $D_F \gg 1$ reflects that across activity, these networks remain highly convoluted with a self-similar structures. For fixed activity and $t_0$ and increasing $\mathcal{T}$, the FTLE fields begin to converge. The inset in Fig.~\ref{fig4}a shows signatures of this, where the rate of change $\mathrm{d}D_F/\mathrm{d}\mathcal{T}$ is found to first increase and then it begins to dip at higher $\mathcal{T}$. Currently, numerical limitations prevent us from probing higher $\mathcal{T}$, which will be explored in more detail in the future. Consistent with the field statistics shown earlier, the backward and forward time ridges yield the same $D_F$. This establishes a further morphological symmetry between attracting and repelling LCS. 

The crossings between forward and backward LCS reflect the local alignment between attracting and repelling fronts, and are linked to flow reversibility~\cite{lapeyre2002characterization}. We quantify the distribution of the angles of the LCS crossings, for which we compute the eigenvectors $\boldsymbol{\xi}_2(\bf x_0)$ of  $\mathbf{C}_{t_0}^{t}(\mathbf{x}_0)$, which give the local \textit{direction} of stretching, along with the magnitude~\cite{haller2015lagrangian}. We find all crossing points and compute the (smaller) angle $\theta$ of intersection between the eigenvectors associated with $\sif$ and $\sib$ ridges. Fig.~\ref{fig4}b shows the distribution of $\theta$, which is found to be close to uniform for all values of activity. This shows that there is no preferential alignment, or orthogonality, between forward and backward LCS. The LCS networks, therefore, consist of isotropic crossings. The inset in Fig.\ref{fig4}b shows the density of crossing points $N_c/N_t$, where $N_c$ and $N_t$ are the number of crossings and total number of points respectively. The crossing density is found to decrease with activity, which hints that there is a change in the relative spatial organization of attracting and repelling LCS, which needs further study. A more detailed analysis of ridge geometry like the distribution of ridge lengths, their curvature and bundling will be pursued in future work.

\section{Conclusions}
In this work, we take the first steps towards uncovering Lagrangian Coherent Structures in active turbulence. Going further, we pitch this study as a comparison of the statistics and geometry of forward and backward time LCS, asking the question whether aspects of Lagrangian irreversibility are manifest in LCS. We find that with increasing activity, changes observed in the flow field viz a multiscale energy distribution, structural hierarchy and changes in flow organization with the emergence of local ordering and heterogeneity, are also reflected in the FTLE fields. In mildly active turbulence these remain densely packed and knotted, reflected in the higher fractal dimension, while for highly active turbulence the network becomes sparse with ridges becoming more elongated and the fractal dimension dipping. Through every metric we have tested, there is a strong symmetry between forward and backward LCS. The FTLE fields are found to have identical distributions and power-spectra, reflecting that both the amplitude and structure of the ridge fields exhibit strong temporal symmetry. Interestingly, despite the flow being vortex dominated as reflected in the skewness of the Okubo-Weiss parameter, strongest mixing both forward and backward in time originates in strain dominated regions. Lastly, the crossings between forward and backward LCS networks are found to be isotropically distributed. 

While there is no existing study outlining a similar comparison in inertial turbulence, we believe this work will spark interest in probing aspects like Lagrangian irreversibility and mixing asymmetry via the lens of LCS across a range of complex flows. From the perspective of active matter, especially in systems like bacterial swarms that can exhibit diverse flowing states, from merely chaotic to intermittent turbulence, depending upon a control parameter like activity (which could be mapped to light intensity or oxygen concentrations in experiments), a deeper understanding of these invariant surfaces forming networks of mixing and transport barriers will greatly aid ongoing efforts to design functionally optimal active matter systems. Such work requires a synthesis between controlling flow reorganization, for instance with spatiotemporal activity modulation to seed or inhibit flow structures, and a deep understanding of mixing and transport mediated by the invariant LCS. Our work opens up this new direction for the taming of active turbulence.

\acknowledgments
SB and KK would like to thank Mattia Serra and Anupam Gupta for the introduction to FTLEs and LCS during the GIAN School on Coherent Structures in Complex Physical and Biological Systems, IIT Hyderabad (2025). We thank Sivasurender Chandran, Dhananjay Gautam and Jason Picardo for discussions. KK acknowledges the TCS Research Fellowship Program (TCS/19/24-25/P49). SM and AM acknowledge the Govt. of India grant ANRF/ECRG/2024/002467/ENS. SM acknowledges the IITK Initiation Grant project IITK/ME/2024316.

\bibliographystyle{eplbib}
\bibliography{references}

\end{document}